\documentclass[twocolumn,showpacs,10pt]{revtex4-1}
\usepackage{amsmath,amssymb,graphicx,upgreek}
\newcommand{\quotes}[1]{``#1''}

\begin{document}

\title{Wide-angle energy-momentum spectroscopy}

\author{Christopher M. Dodson}
\altaffiliation{These authors contributed equally to this work.}

\author{Jonathan A. Kurvits}
\altaffiliation{These authors contributed equally to this work.}

\author{Dongfang Li}
\author{Rashid Zia}
\altaffiliation{Corresponding author: rashid\_zia@brown.edu}

\affiliation{School of Engineering and Department of Physics, Brown University, Providence, RI 02912}

\begin{abstract}Light emission is defined by its distribution in energy, momentum, and polarization. Here, we demonstrate a method that resolves these distributions by means of wide-angle energy-momentum spectroscopy. Specifically, we image the back focal plane of a microscope objective through a Wollaston prism to obtain polarized Fourier-space momentum distributions, and disperse these two-dimensional radiation patterns through an imaging spectrograph without an entrance slit. The resulting measurements represent a convolution of individual radiation patterns at adjacent wavelengths, which can be readily deconvolved using any well-defined basis for light emission. As an illustrative example, we use this technique with the multipole basis to quantify the intrinsic emission rates for electric and magnetic dipole transitions in europium-doped yttrium oxide (Eu$^{3+}$:Y$_{2}$O$_{3}$) and chromium-doped magnesium oxide (Cr$^{3+}$:MgO). Once extracted, these rates allow us to reconstruct the full, polarized, two-dimensional radiation patterns at each wavelength.
\end{abstract}

\maketitle 
When light is emitted by an electronic system, it radiates into the optical modes of its local environment with a characteristic distribution in energy, momentum, and polarization. These distributions or spectra can reveal a tremendous amount of information about the electronic structure of the emitter as well as its optical environment. For example, Fourier-space measurements of radiation patterns can be used to identify the orientation of single fluorescent molecules~\cite{LiebJOSAB2004}, study how quantum dots interact with optical antennas~\cite{CurtoScience2010}, and quantify the multipolar origin of electronic transitions in solid state ions~\cite{TaminiauNatCom2012}. As researchers seek new ways to direct and enhance the radiation from single emitters for applications in nanophotonics and quantum optics, there is a growing need for techniques that can extract maximal information from low-light samples.

In conventional spectroscopy, one often seeks to maximize the signal from one distribution by either integrating over or filtering out the other two. For example, to measure the energy (i.e. wavelength) spectrum of weakly luminescent samples, one typically integrates over all collection angles and polarizations. Although this maximizes the signal in the energy spectrum, it does so at the expense of eliminating information contained in the other two distributions.

Nevertheless, measuring momentum and polarization information has become increasingly important in nano-optics. For example, two-dimensional radiation patterns have helped verify directional emission from quantum emitters coupled to optical antennas and photonic crystals~\cite{CurtoScience2010,WagnerAPL2012,ZhuNanoLett2012,WangNanoLett2013,CurtoNatComm2013,HancuNanoLett2013,Wu2DMat2014}. To obtain such images, researchers have used either bandpass filters to study emission at specific wavelengths ~\cite{CurtoScience2010,CurtoNatComm2013,HancuNanoLett2013} or scanning systems to obtain spectral information at select angular cross-sections ~\cite{WagnerAPL2012,ZhuNanoLett2012,WangNanoLett2013}. However, using these techniques to obtain a complete analysis of the spectral and angular emission would substantially increase the measurement time and/or require an impractical number of bandpass filters.  As such methods inherently reduce the amount of collected light, they cannot readily scale to produce high resolution measurements in all dimensions.  

\begin{figure}[b]
\centerline{\includegraphics[scale=1]{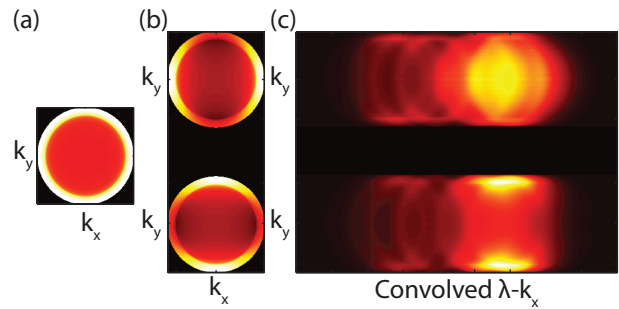}}
\caption{Wide-angle energy-momentum spectroscopy method. Momentum-space emission pattern (a) is separated by Wollaston prism into orthogonal polarizations (b) and dispersed by a grating into convolved energy-momentum spectra (c).}
\label{fig:ConvolvedSpectrum}
\end{figure}
Here, we demonstrate a spectroscopic imaging technique that maximizes the total luminescent signal while also retaining information about the polarized angular (momentum) and spectral distributions. To this end, we image the back focal plane of a microscope objective using a Bertrand lens to obtain a Fourier-space radiation pattern, Fig.~\ref{fig:ConvolvedSpectrum}(a). With a Wollaston prism, this pattern is projected onto two spatially separated and orthogonal polarizations, Fig.~\ref{fig:ConvolvedSpectrum}(b).  Then, using a diffraction grating, these two-dimensional emission patterns are dispersed and reimaged, Fig.~\ref{fig:ConvolvedSpectrum}(c). This approach builds on recent work ~\cite{TaminiauNatCom2012,SchullerNatNano2013,KaraveliNano2013}, hereafter refered to as narrow-slit energy-momentum spectroscopy, but eliminates the need to filter polarization and momentum information using a linear polarizer and entrance slit. Wide-angle energy-momentum spectroscopy thus convolves momentum and spectral information. In this letter, we present a method to deconvolve these overlapping distributions using the well-defined multipolar basis, and exploit this method to reconstruct the polarized, momentum distribution at each wavelength.

As suggested by recurring annular shapes, the spectrally dispersed distribution of light in Fig. 1(c) retains information about the underlying emission patterns. In order to resolve these 2D radiation patterns, we first examine the convolution process. The wide-angle energy-momentum spectra can be defined as a superposition of wavelength-dependent radiation patterns from different basis functions.  For the specific case of isotropic electric dipole (ED) and magnetic dipole (MD) transitions, Fig. 2(a) shows how the dispersed image is a superposition of ED and MD radiation patterns centered at different wavelengths. These basis functions are analytically derived from the Green's Dyadic functions for the electric and magnetic fields in a layered dielectric ~\cite{TaminiauNatCom2012, NovotnyBook}. In this context, the local density of optical states, as defined by the sample's optical structure, encodes the possible distributions for light emission. As a result, the number of free parameters (typically $\sim10^3$ $x_{m,\lambda_i}$ coefficients) is much smaller than the number of measurement values (typically $10^5\sim10^6$ pixels in the image data, $b$).

To deconvolve this highly overdetermined system, we construct a sparse measurement matrix, $A$, as shown in Fig.~2(b). Each column of $A$ contains the wavelength and polarization specific radiation patterns associated with a single basis function. Note that these columns have been offset to account for the dispersion introduced by the grating. The non-zero elements in each row of $A$ describe the overlapping terms (from different wavelengths or basis functions) that may contribute to light detected at a specific pixel. In general, the measurement matrix could be expanded to include other operations and optical components (e.g. waveplates, filters, or electrical readout noise). For simplicity though, we restrict $A$ to consist of the basis functions alone and introduce, as a simple offset, a single constant background term, $\eta$. Since the number of measurement values in $b$ is two to three orders of magnitude larger than the number of free parameters in $x$, we employ a ridge regression method to solve a convex optimization program of the form:\begin{eqnarray}
\text{minimize } &\|Ax&-b+\eta\|_2 + \alpha\|\Delta x\|_2\label{Eq:CvxRoutine}\\ 
\text{subject to} &x &\geq{}0\nonumber. 
\end{eqnarray}

\begin{figure}[t]
\centerline{\includegraphics[scale=1]{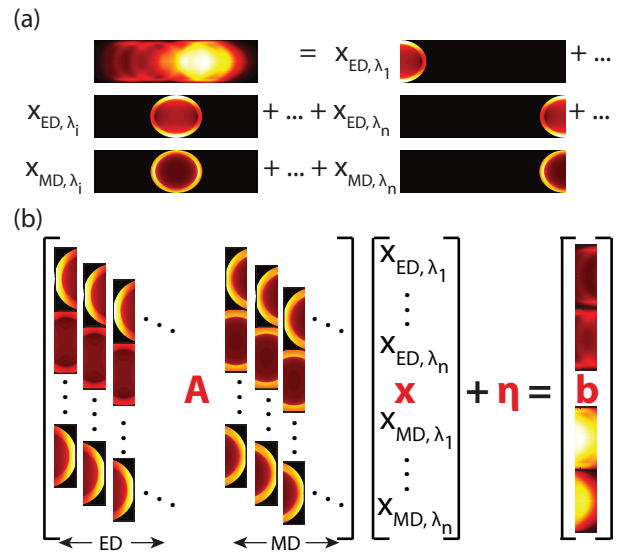}}
\caption{Convolution model and decomposition method (shown for simplicity with single polarization). (a) Convolved wide-angle energy-momentum data can be decomposed into a superposition of radiation patterns, where $x$ coefficients describe the contribution of basis functions (e.g. ED and MD) at each wavelength, $\lambda_i$.  (b) Schematic representation of mathematical model used for decomposition process highlighting how basis functions and image data are vectorized in $A$ and $b$ to form a matrix equation.}
\label{fig:MeasurementMatrixSchematic}
\end{figure}

Here, $x$ is a vector argument defining the contribution of each basis function to light emission at each wavelength. For the present multipolar basis, $x$ is equivalent to the intrinsic emission rates (i.e. spectrally resolved Einstein $A$ coefficients) in \cite{TaminiauNatCom2012}. The vectorized measurement data, $b$, is obtained from the CCD detector. The parameter $\alpha$ defines the penalty for variations in $x$, as described by the $l^2$-norm of the finite difference $\Delta{}x\equiv x_i - x_{i-1}$. By letting $\alpha$~=~0, Eq.~(\ref{Eq:CvxRoutine}) would simplify to a non-negative constrained least squares problem, but in this paper, we choose $\alpha$~=~0.025 to promote continuity in the intrinsic rates, i.e. penalize high frequency variations~\cite{Note1}. To fit $x$, we use the CVX convex optimization package for Matlab with the MOSEK solver ~\cite{CVX,GrantCVX2008}.

\begin{figure*}
\centerline{\includegraphics[scale=1]{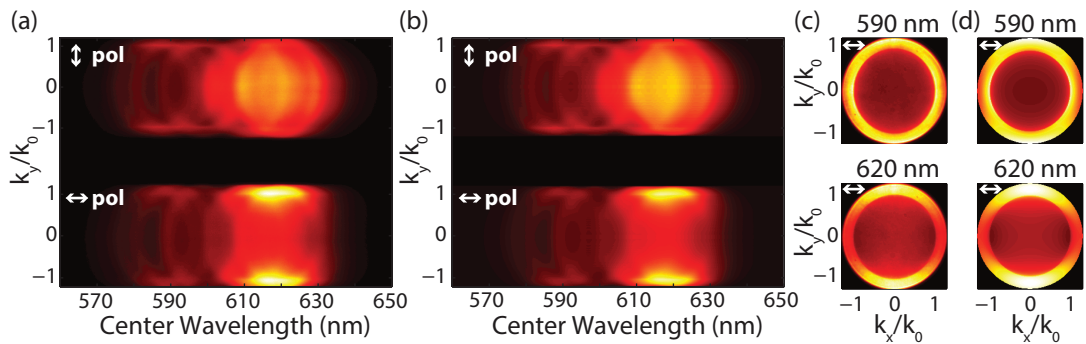}}
\caption{Wide-angle energy-momentum spectroscopy of Eu$^{3+}$:Y$_2$O$_3$. (a) Experimental wide-angle data and (b) corresponding theoretical fits obtained using Eq.~\ref{Eq:CvxRoutine}. (c) Experimental polarized 2D radiation patterns using 590$\pm$10 and 620$\pm$15~nm bandpass filters and (d) corresponding 2D radiation patterns extracted from wide-angle fits. White arrows denote polarization.}\label{Figure3}
\end{figure*}

To demonstrate this technique, we revisit the emission from a 20 nm thin film of Eu$^{3+}$:Y$_2$O$_3$ that was originally characterized using the narrow-slit technique~\cite{TaminiauNatCom2012}. Wide-angle energy-momentum spectra were obtained by exciting Eu$^{3+}$ ions with a 532 nm pump laser focused on the sample through a 1.3 NA 100x (Nikon Plan Fluor) oil immersion objective mounted on an inverted microscope. A field stop was placed at the exit port of the microscope, before the Wollaston prism, to suppress background fluorescence. A 100 mm focal length Bertrand lens was then used to image the back focal plane of the objective onto the open entrance port of a Schmidt-Czerny-Turner (SCT) imaging spectrograph (Princeton Instruments IsoPlane). After dispersal by a diffraction grating (300 lines/mm blazed at 750 nm), energy-momentum spectra were imaged using a CCD camera (Princeton Instruments Pixis 1024B).

Figure~\ref{Figure3}(a) shows the experimental wide-angle energy-momentum data. Figure~\ref{Figure3}(b) shows theoretical fits obtained by solving Eq.~(\ref{Eq:CvxRoutine}) using the isotropic ED and MD basis. The good agreement between theory and experiment is substantiated by a low normalized root-mean-square error (NRMSE) of 4\%. Having deconvolved the wide-angle energy-momentum spectra, we can reconstruct individual radiation patterns in any desired wavelength range.  In Fig.~\ref{Figure3}(c), we show experimental radiation patterns obtained using two different bandpass filters, 590$\pm$10 and 620$\pm$15 nm respectively. In Fig.~\ref{Figure3}(d), we show the corresponding radiation patterns extracted from the wide-angle fits, and again observe good agreement (6\% NRMSE). While only two examples are shown here, this method simultaneously extracts 2D radiation patterns at each and every wavelength within the measurement domain from a single acquisition.

To further validate these results, Fig. 4(a) compares the intrinsic emission rates extracted from the wide-angle measurements to momentum-filtered measurements made with the inclusion of a $\sim$10$~\upmu$m slit at the entrance port of the spectrograph. The normalized intrinsic emission rates display good qualitative agreement with each other as well as previous measurements~\cite{TaminiauNatCom2012}. More importantly, the two methods yield quantitatively similar extracted emission rates, as evidenced by a low NRMSE of 3\%. This suggests that convolution and deconvolution do not significantly deteriorate the extracted information. While the wide-angle technique exhibits greater variation in the rates, it allows for a ${>}10^2$ fold increase in collected light. For example, the narrow-slit measurements required an acquisition time of 500 seconds, whereas the wide-angle measurements were performed in 10 seconds and still yielded over 2.2 times more counts (i.e. a 110x improvement in optical throughput). 

\begin{figure*}
\centerline{\includegraphics[scale=1]{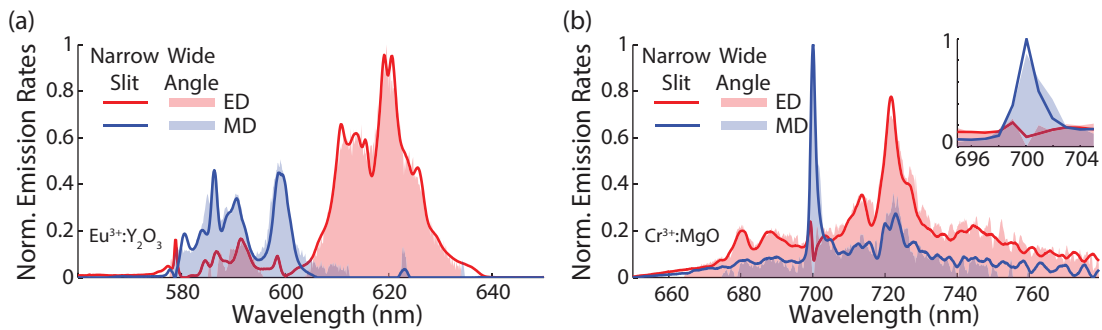}}
\caption{Comparison of narrow-slit and wide-angle energy-momentum spectroscopy for (a) Eu$^{3+}$:Y$_2$O$_3$ and (b) Cr$^{3+}$:MgO. Extracted ED (red) and MD (blue) normalized emission rates from measurements both with (line) and without (shaded area) a $\sim$10$~\upmu$m entrance slit. The inset of (b) highlights the inverted symmetry of ED and MD emission at the ZPL near 700 nm.}
\label{Figure4}
\end{figure*}

In addition to improving acquisition time, this wide-angle method can also improve the effective spectral resolution. To demonstrate this point, we study the narrow linewidth zero phonon line (ZPL) in Cr$^{3+}$:MgO, which is ED forbidden and MD allowed. The same measurement setup as described above was used to obtain both wide-angle and narrow-slit energy-momentum spectra from a 45 nm Cr$^{3+}$:MgO thin film deposited on a quartz substrate via electron beam evaporation. The resulting extracted intrinsic rates are shown in Fig.~\ref{Figure4}(b). Here, we see that the ZPL near 700 nm corresponds to a peak in the MD intrinsic rates that is correlated to a sharp decrease in the ED intrinsic rates. While this feature is visible in the narrow-slit results, the ED dip is more pronounced in the wide-angle measurements, nearly going to zero as one would expect for an ED forbidden line. Interestingly, by removing the entrance slit, the wide-angle measurements do not necessarily sacrifice resolution; since we are looking at the distribution of light around each wavelength, the wide-angle deconvolution can still resolve features at the single pixel limit. Conversely, when working with weakly luminescent emitters with conventional slit-based spectrographs, one is often forced to enlarge the entrance slit in order to acquire sufficient emitted light at a reasonable signal-to-noise ratio. This results in an averaging effect over nearby wavelengths, which can smooth out the emission spectrum and remove sharp features. In fact, previous studies of multipolar emission in Cr$^{3+}$:MgO~\cite{KaraveliNano2013} did not exhibit the inverted symmetry of the ED and MD rates at the ZPL shown here. Thus, despite removing the slit on the entrance port of the spectrograph, we are not only increasing the \'{e}tendue of the optical system, but also improving the effective spectral resolution for practical measurement constraints.

In conclusion, we have presented a spectroscopic technique that allows for the simultaneous measurement of the momentum and spectral distribution of light emission, while also maximizing the collected signal. This technique decreases the need for multiple measurements to obtain each of these distributions, and also ensures that the \'{e}tendue of the optical system is only limited by the collection efficiency of the microscope objective. Despite convolving the momentum and spectral distributions, we were able to extract intrinsic emission rates and recover spectrally dependent radiation patterns, because the emitted light within a thin film sample can be decomposed into the multipolar basis.  More generally, this technique can be used to resolve the emission or scattering of light in any system for which the local density of optical states can be either analytically or numerically defined. Similar to the physical masks used to encode information in Hadamard spectroscopy ~\cite{MendeApplOpt1993,HarwitBook1979}, the local density of optical states in a structured environment encodes a well-defined pattern onto light emission, which allows one to maximize throughput without sacrificing information or resolution. By including additional polarization projections, this technique could also be extended to obtain the entire polarization distribution~\cite{PerreaultOptLett2013}. We anticipate that wide-angle energy momentum spectroscopy will be particularly helpful in characterizing the dispersive and highly asymmetric radiation patterns for directional antennas~\cite{CurtoScience2010,ZhuNanoLett2012,WangNanoLett2013,CurtoNatComm2013,HancuNanoLett2013}, oriented emitters~\cite{LiebJOSAB2004,SchullerNatNano2013}, photonic crystals~\cite{WagnerAPL2012,Wu2DMat2014}, and other nanophotonic systems. 

The authors thank M. Jiang for helpful discussions. This work was supported by the Air Force Office of Scientific Research (FA9550-09-1-0346, PECASE FA9550-10-1-0026, and MURI FA9550-12-1-0488) and the National Science Foundation (CAREER EECS-0846466).

\end{document}